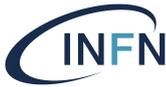

ISTITUTO NAZIONALE DI FISICA NUCLEARE

# THE SCIENTIST' EXPERIENCE IN PARTICIPATED SCIENCE COMMUNICATION

G. Mazzitelli[1,2], P. Bolaffio[2,3], G. Burzachechi[2,3], I. Capra[2], G. Ciocca[2,4], A. Della Ceca[2,4], R. Giovanditti[2], C. Grasso[2], D. Maselli[1,2], G. Sanzone[2], F. Spagnoli[2], M. Tota[1,2]

[1]Istituto Nazionale di Fisica Nucleare Laboratori Nazionali di Frascati
Via E. Fermi 00044 Frascati, Rome, ITALY

[2]Associazione Frascati Scienza, Scuderie Aldobrandini,
Piazza Marconi 3 – 00044 Frascati, Rome, ITALY

[3]Ass. prom. Soc. Il Refuso,
Via Battaglia di Pontegrande 7b. 00040 Monte Porzio Catone (Rm)

[4]G.Eco - Gruppo didattica dell'Ecologia

Piazza delle Iris, 28, 00171, Rome, ITALY

## Abstract

Since 2006 a small group of researchers from the Italian National Institute for Nuclear Physics started to realized one of the first European Researchers' Night in Europe: a one night-event, supported by the European Commission, that falls every last Friday of September to promote the researcher's figure and its work. Today, after thirteen editions, the project has evolved by involving more than 60 scientific partners and more than 400 events/year spread from the North to the South of Italy in 30 cities, captivating more than 50.000 attendees with a not negligible impact on the people and the press. During the years, the project has followed and sometimes anticipated the science communication trend, and BEES (BE a citizEn Scientist) is the last step of this long and thrilling evolution that brought to a huge public engagement in our territory.

The experience, the methodology, and the major successful examples of the organized events are presented together with the results of the long term project impact.

## 1 INTRODUCTION

The European Researchers' Nights (NIGHT) [1] are public events, spread all over Europe, dedicated to bring researchers closer to the public. The objective of the European project is to highlight the impact of research on our daily lives and to motivate young people to embark on research careers. Last year 55 projects have been implemented in 371 cities across Europe and beyond and over 1.5 million visitors attended. Frascati, a small town 30 km far from Rome, is the epicenter of a research area where 3000 scientists are working, where there are 8 research institutes, very close to many universities where young students and researchers came from.

In 2006 a small group of researchers from the Italian National Institute for Nuclear Physics started to realize one of the first NIGHT [2], an experience that continues up to nowadays with the preparation of the 14th edition. During the years many institutions and cities spread all over Italy have joint the project today carried on by the Frascati Scienza Association [3]. The NIGHT projects not only successfully pursued the objectives of the European call, but also worked to make a real change in the perception of the general public, stakeholders and researchers of what research and science communication is, always innovating its methods and messages. Starting from the objective to make accessible large research infrastructures (COME IN 2006), up to showing science as the driver for creating effective benefits to the European citizens (MADE IN SCIENCE 2017), Frascati Scienza has made a real impact on the territory, scientific partners and institutions, as shown by the impact assessment results of the past 13 years, as well as increasing constantly the number of attendees (today more than 50.000) and the quality of the organization and partnership process (see figure 1).

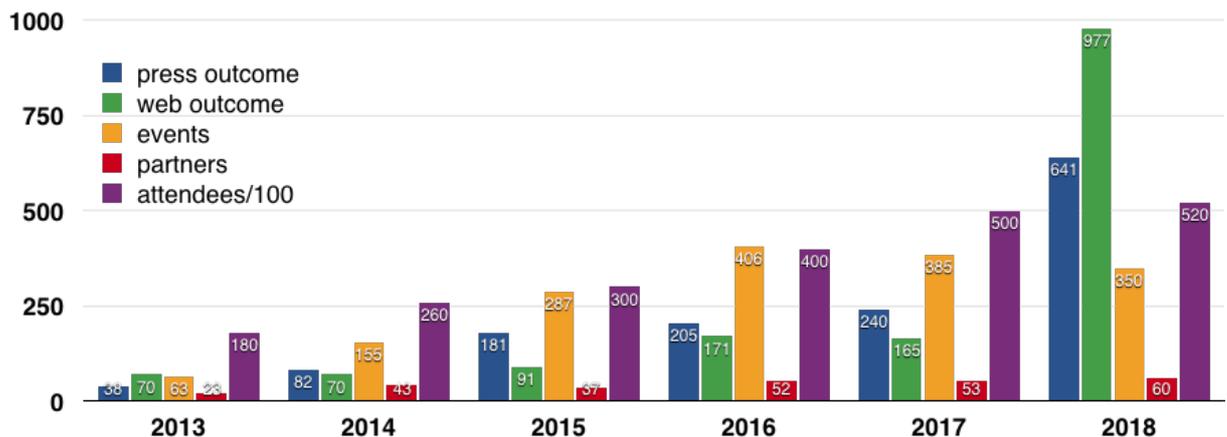

*Figure 1. Summary of some important indicators of recent Frascati Scienza's NIGHTs: the **press outcome** is the number of accredited news outputs (press, tv, radio, etc) triggered by Frascati Scienza NIGHT press releases; **web outcome** is the number of websites, blogs, etc speaking about Frascati Scienza NIGHT; **events** are the number of events spread over the cities and **partners** operating with Frascati Scienza; **attendees/100** is the number of "hecto" participants in the different Frascati Scienza's NIGHTs.*

Frascati Scienza, designing and developing the NIGHT for this last two years (2018-2019), believed that a further step should be done in the direction of science, and researchers' work in general, to actively and fully participate and collaborated by the society. This requires, for this project and more in general for science communication, to enact a new paradigm in which research do not belong only to researchers but come out from a common collaboration process with the society and the various stakeholders to enjoy science discoveries.

For these reasons in the implementation of its activities, this edition of the NIGHTs (2018-2019) start to directly involve the citizens, and especially young people, in hands-on activities, such as "**mini**" **citizen science** projects, by collecting and using data sets, encompassing different disciplines, with the ultimate goal to acquire new knowledge, scientific skills and competencies, supporting the public recognition of researchers. This process demonstrates how science proceeds by trials, but comes to solid conclusions and incontrovertible results, so people can trust the hard work of researchers.

## 2 METHODOLOGY

*2.1 BE a CitizEn Scientist methodology*

The best expression of a participating model where people can take an active part in discussions, games and research itself is the "citizen science" activity, where the single user can help researchers in their scientific projects.

Frascati Scienza promoted the design and the implementation of specific scientific activities and projects in medium and long term, but also in one-day-events, in order to involve the public on exploring and experimenting with science, taking an active part in discussions, games and research itself. The experiments are completely managed by the general public, along with the support of the researchers, taking advantage of ICT technologies for collecting data, enabling society to directly analyse this information and creating new scientific knowledge.

Approximately 400 events, designed for the general public, are mainly based on science shows, hands-on experiments, demos, simulations, workshops, games, using the methodology summarized as follows:

- Citizen Scientists, coding classes, role games, robotics and other interactive experiences.
- GigaLabs, great science games designed especially for kids and teens with a high impact scenography.
- School of Science, science classes inside the schools, led by researchers and scientific explainers.
- Be a part of the Researchers' Night, young students selected helping the organization of the events, in order to enable them to understand how to plan and realize this kind of activities.
- Research internship, many high school students had the possibility of working inside several Italian Research Centers.
- Guided tours Inside research centers, inside all the Research Institutes and Universities involved in project, several interactive activities have been planned (such as Science shows, workshop for kids and teens, interactive exhibitions etc.).
- Science Trips Target, bring research (and researchers) outside the laboratories and at the same time to valorize the huge cultural heritage of our Country.
- Street Scientists, the researchers in the streets and squares of the cities offer their time and knowledge to answer questions and curiosities of the public.
- Science Chats Target, informal scientific activities, such as science cafés, science aperitifs and books' presentations, organized inside pubs, bars and libraries of the cities involved in the project so that people will have the possibility to debate with researchers about common life topics, ask for suggestions about books or music in a friendly way, as well as to discuss scientific topics.
- Science Shows, fun and interactive events to talk about the scientific challenges of the future, and to better understand how citizens and society can participate in science and research, taking into account he ludic and spectacular aspects of science communication are taken.
- Science Exhibitions, science exhibitions, planned to be interactive through various media, with the participation of a researcher or a communicator.

Moreover, one example of "mini" citizen science project is Be a Data Scientist, developed to investigate which are the dynamics of information among young and very young people.

*2.1.1 Be a Data Scientist*

Be a Data scientist has been divided into five-phase:

Phase 1: some schools class, defined in the following as working groups (WG), has been selected around Italy and a scientific committee has been set up. The members of the scientific committee were chosen from scientific journalists, expert of social media, data scientists, experts of demographic statistics.

Phase 2: each WG has proposed its own survey to investigate the questions proposed by the scientific committee.

Phase 3: the scientific committee has re-elaborated the questionnaire, and has sent it to the WG for the administration to colleagues, parents, etc. through a google web app for data collection.

Phase 4. Each WG is now going to analyze data under the supervision of scientists and teachers (coding, statistics, machine learning).

Phase 5: Each WG will make its results processing and discusses the results with the scientific committee, and will present the results at the next edition of the European Researchers Night.

The project collected 1021 valid survey and now the students are going to start analyzing data. This phase expected the usage of coding based on python and shared on "colab research" by google [4].

# 3 RESULTS

*3.1 BE a CitizEn Scientist results*

The analysis of the impact of the European Researchers' Night was delivered at quantitative (questionnaire) and qualitative level (Trivia Night event) [5] .

*3.1.1 Quantitative impact*

The total of questionnaires collected is 2029 (786 Ex-Ante, 1074 Ex-Post and 169 Kids).

Without any significant variation compared to the information previously analysed in the Ex-ante questionnaire, most of the Ex-Post respondents thinks that the ERN 2018 has contributed to let people understand that science is fun (60% responded a lot), and that the event is a good way to have citizens actively involved in science (57% responded a lot).

The majority of the respondents to the ex-post questionnaire are female (61%), in the age range of 30-39 years old, followed by 10-19 and 40-49 years old people. Most of them have a Master Degree (29%) and/or a High School Degree (29%), followed by a Primary School Degree (19%).

Most of the respondents to the ex-post questionnaire knew about the ERN 2018 thanks to friends (35%) and websites (34%), while a consistent part of them responded through Facebook (20%).

More than half of the respondents to the ex-post questionnaire of ERN 2018 has already participated in previous editions of the event.

The majority of them think that the ERN 2018 has strongly contributed to show mainly that research is fun (60% replied a lot) and why it is important doing research (54% replied a lot). Mostly all of the participants agree that the role of research is important for both the development of our country (73% replied a lot) and of Europe (73%).

*3.1.2 Qualitative impact*

The Trivia Night event is a quiz by which the staff of Frascati Scienza involved the general public to gather impressions and suggestions about the event, providing their qualitative opinion about the organization and results achieved in the ERN 2018. This year, the questions were focused on the citizen science, in order to understand how much people are familiar with this concept, and as a result to increase their knowledge about it.

We presented 7 projects funded at international level by the European Commission, selected to be the most successful ones in terms of development of citizen science activities. For each project, after a short introduction, we challenged the participants to develop a scientific test to define some results.

Among the citizen science projects presented during the Trivia Night, the most interesting to the people was the one that supports kids to become the scientists of the future.

The main information about the evaluation of ERN 2018 resulted from specific questions about that, such as:

- *Why the researchers' night is useful?* Most of the respondents said that the event is useful to let people understand that research is fun, to learn new things and enable the society to be part of science. It was interesting to see that they think the event is useful to not commit the same mistakes of the past and to prepare kids for the future. Moreover, the European Researchers Night raises awareness on issues relevant for the whole community, not only on science.

- *Define 5 motivations about why the ERN 2018 has been successful.* Because is for everyone. Contents include a lot of discipline. It is very interactive. Because of the staff and their work. For the gadgets. Events' organization, knowledge, inspire, fun. Very well advertised, stimulates the mind.

*3.2 Be a Data Scientist results*

A preliminary visualization of the data collected for the Be a Data Scientist project already shows some nice correlation about the dynamics followed by youngsters of how they communicate on media and spread the news. The 1021 samples are composed of 728 (56% women) surveys provided by youngsters under 25, mainly below 17 years old, and 293 (82% woman) surveys provided by people with more than 25, 45 years old on average. Firstly, while 74% of adults daily search for news, only 45% of young people do that. In order to be informed, youngsters do not use printed papers and radio as adults do, while the internet and TV look to be the main channels.

Concerning news scaring people, youngs and adults look to be similar apart for a large concern of adults about violence and child abuse. While some differences are appearing about how often youngsters and adults look for scientific news: only 10% of youngsters against 30% of adults do a search of scientific news daily, and youngsters don't like the question mainly no answering. In the case of dramatic events, the youngsters search videos, while adults search photos and comments. Anyway, the main source of information for these dramatic events, especially for young people, is the TV, not the network.

About the online news, the young prefer to obtain news from social networks and browsers, and their preferred channel is Instagram followed by Youtube, while for adults it is Facebook. While adults are mainly attracted by the contents of the news, young confess to be attracted mainly from photos or memes and videos related. Both are attracted by chronicle and latest news, but the young are also strongly attracted by sports, music and TV series. Relatives and friends are the main targets of re-posting for both the sample. Moreover, for the young, the main channel of sharing it is to talk directly with friends, followed by WhatsApp and sms. The young repost more than adults because impressed by the news, and only 2% do that to imitate others. 42% of the young (65% among adults) think to know what is digital reputation, and 84% (91% among adults) know which hazards are involved when posting photos on the internet.

Data can be downloaded and visualize on the gitlab repository [6].

## 4   CONCLUSIONS

The European researchers' nights were, during these 14 years, of fundamental importance for the success of scientific communication, the reputation of researchers and the approach to the scientific careers of young people. In particular, the Frascati Scienza' NIGHTs have been a positive impact on the territory that today is strongly demanding of taking part in science. In this framework, the idea to introduce mini citizen science project has been successful in terms of societal impact and communication as demonstrated by the huge feedback on media, social and web outcomes. The mini citizen science project Be a Data Scientist is a successful example of this approach in which Frascati Scienza is applying a participative model to involve young and people at large in science.

## ACKNOWLEDGEMENTS

We would like to thank Colette Renier, P.O. of the NIGHT since 2005, recently retired, for the inexhaustible support during these years, the thousands of researchers involved and the institutions, associations, and companies supporting and founding the manifestation [7].

Work supported by EU MSCA-NIGHT-2018/2019 - Grant Agreement No. 818728